\documentclass{sig-alternate}
\usepackage{enumitem}
\usepackage{times}
\usepackage{latexsym}
\usepackage{url}
\usepackage{color}
\usepackage{multirow}
\usepackage{comment}
\usepackage{amsmath} 

\usepackage{amssymb}

\usepackage{etoolbox}

\usepackage{graphicx}
 
 \usepackage{enumitem}
  {\begin{itemize}[topsep=0pt, partopsep=0pt] \footnotesize%
    \setlength{\itemsep}{0pt}%
    \setlength{\parskip}{0pt}%
    }%
  {\end{itemize}}

  {\begin{itemize}[topsep=0pt, partopsep=0pt]%
    \setlength{\itemsep}{0pt}%
    \setlength{\parskip}{0pt}%
    }%
  {\end{itemize}}  
\usepackage{multirow}

\newenvironment{tightenu}%
  {\begin{enumerate}[topsep=0pt, partopsep=0pt] \footnotesize%
    \setlength{\itemsep}{0pt}%
    \setlength{\parskip}{0pt}%
    }%
  {\end{enumerate}}

\newcommand\djc[1]{\textcolor{red}{[DJ: #1]}}

\begin{document}
\setlist{nolistsep}

\title{Inferring User Preferences by Probabilistic Logical Reasoning over Social Networks}
%
%
%
%
%

\numberofauthors{8} 
%
\author{
%
%
\alignauthor
Jiwei Li\\
       \affaddr{Computer Science Department}\\
       \affaddr{Stanford University}\\
       \affaddr{CA, USA, 94306}\\
       \affaddr{jiweil@stanford.edu}
\alignauthor
Alan Ritter\\
       \affaddr{Dept. of Computer Science and Engineering}\\
       \affaddr{The Ohio State University}\\
       \affaddr{OH, USA, 43210}\\
       \affaddr{Ritter.1492@osu.edu}
\alignauthor{Dan Jurafsky}\\
       \affaddr{Computer Science Department}\\
       \affaddr{Stanford University}\\
       \affaddr{CA, USA, 94306}\\
       \affaddr{jurafsky@stanford.edu}
}
\date{30 July 1999}

\maketitle
\begin{abstract}
We propose a framework for inferring the latent attitudes or preferences of users 
by performing probabilistic first-order logical reasoning over the social network graph.
Our method answers questions about Twitter users like {\em Does this user like sushi?}
or {\em Is this user a New York Knicks fan?} by building a probabilistic model that reasons over
user attributes (the user's location or gender) and the social network (the user's friends and spouse),
via inferences like homophily (I am more likely to like sushi if spouse or friends like sushi,
I am more likely to like the Knicks if I live in New York).
The algorithm uses distant supervision, semi-supervised data harvesting and vector space models 
to extract user attributes (e.g. spouse, education, location) and preferences (likes and dislikes) from text.
The extracted propositions are then  fed into a probabilistic reasoner (we investigate both
Markov Logic and Probabilistic Soft Logic).
Our experiments show that probabilistic logical reasoning significantly improves the performance
on attribute and relation extraction, and also achieves an F-score of 0.791
at predicting a users likes or dislikes, significantly better than two strong baselines.
\end{abstract}

\category{H.0}{Information Systems}{General}

\keywords{Logical Reasoning, User Attribute Inference, Social Networks} 

\section{Introduction}

Extracting the latent attitudes or preferences of users on the web is an important goal,
both for practical applications like product recommendation, 
targeted online advertising, friend recommendation, or for 
helping social scientists and political analysts gain insights into public opinion and user behavior.

Evidence for latent preferences can come both from
attributes of a user or from preferences of other people in their social network.
For example people from Illinois may be more likely to like the Chicago bears,
while people whose friends like sushi may be more likely to like sushi.

A popular approach to  draw on such knowledge to help
extract user preferences is to make use of collaborative filtering,
typically applied on structured data describing explicitly provided user preferences (e.g. movie ratings),
and often enriched by information from a social network
\cite{debnath2008feature,goeksel2010system,kautz1997referral,konstas2009social,liu2010use}.
These methods can thus combine information from shared preferences and attributes with
information about social relations.

In many domains, however, these  user preferences and user attributes 
are not explictly provided, and we may not even have explicit knowledge of relations
in the social network.  In such cases, we may first need to estimate these latent attributes or preferences,
resulting in only probabilistic estimates
of each of these sources of knowledge.  How can we reason about user preferences
given only weak probabilistic sources of knowledge about users' attributes,
preferences, and social ties? This problem occurs in domains like Twitter,
where knowledge about users' attitudes, attributes,  and social relations must
be inferred.

We propose to infer user preferences on domains like Twitter without explicit information
by applying relational reasoning frameworks  like
Markov Logic Networks (MLN) \cite{richardson2006markov} and Probabilistic Soft Logic  (PSL) \cite{goertzel2014probabilistic}
to help  infer these relational rules.
Such probabilistic logical  systems are able to combine evidence probabilistically to draw 
logical inference.

For example, such systems could learn individual probabilistic inference rules expressing facts like
``people who work in IT companies like electronic devices" with an associated probability (in this case 0.242):

\begin{quote}
\textsc{Work-In-IT-company(A)}$\Rightarrow$ \textsc{like-electronic-device}  (0.242)
\end{quote}

Such systems are also able to perform global inference over an entire network
of such rules, combining probabilistic information about user attributes, preference, and relations
to predict preferences of other users.

Our algorithm has two stages.  In the first stage we 
extract user attributes.
Unlike structured knowledge bases such as Freebase and Wikipedia,
propositional knowledge describing the attributes of entities in social networks is very sparse.
Although some social media websites (such as Facebook or LinkedIn) do support structured data format of personal attributes,
these attributes may still be sparse, since only a small proportion of users fill out any particular profile fact,
and many sites (such as Twitter) do not provide them at all.
%
On the other hand, users of online social media frequently publish messages describing their preferences and activities, often explicitly mentioning attributes 
such as their \textsc{Job, Religion}, or {\sc Education} \cite{li2014weakly}.  
We propose a text extraction system for Twitter that 
combines supervision \cite{craven1999constructing}, semi-supervised data 
harvesting (e.g., \cite{kozareva2010learning,kozareva2010not}) and vector space models \cite{bengio2006neural,mikolov2010recurrent}
to automatically extract structured profiles from the text of users' messages. 
Based on this approach, we are able to construct a comprehensive list of personal attributes which are explicitly mentioned in text (e.g., \textsc{Like/Dislike, Live-in, Work-for}) and user relations  (e.g., \textsc{Friend,Couple}).

While coverage of user profile information can dramatically be increased by extracting information from text, not all users explicitly mention all of their attributes.
To address this, in the second stage of our work we further investigate whether it is possible to extend the coverage of extracted user profiles by inferring attributes not explicitly mentioned in text through logical inference.  

Finally, we feed the extracted attributes and relations 
into relational reasoning frameworks, including Markov Logic Networks (MLN) \cite{richardson2006markov} and Probabilistic Soft Logic  (PSL) \cite{goertzel2014probabilistic},
to infer the relational rules among users, attributes and user relations that allow us to
predict user preferences.


We evaluate the
system on a range of prediction tasks including preference prediction (liking or disliking)
but also attributes like location or relations like friend-of.

The system described in this paper provides new perspectives 
for understanding, predicting interests, tendencies and behaviors of social media users in everyday life.
While our experiments are limited to one dataset, Twitter, the techniques are
general and can be easily adapted with minor adjustments. 
The major contributions of this paper can be summarized as follows:
\begin{itemize}
\item We present an attempt to perform probabilistic logical reasoning on social networks.
\item Our  framework estimates the attributes of online social media users without  requiring explicit mentions.
\item Our  framework combines probabilistic information about user attributes, preferences, and relations
to predict latent relations and preferences.
\item  We present a large-scale user dataset specific for this task.
\end{itemize}

The next sections show
how user attributes, relations, and preferences are extracted from text and introduce
the probabilistic logical frameworks.  Our 
algorithm and results are illustrated in Section 4 and 5.

\section{Extracting Probabilistic Logical Predicates}
\label{extraction}

Given the message streams from Twitter users, our first task is to extract
information about user attributes, relations, and preferences in logical form;
these will then be input to our global logical inference network.

We represent these facts by two kinds of propositional logic objects: {\bf predicates} and {\bf functions}. 
Functions represent mappings from an object to another object, returning an object, as in \textsc{CapitalOf(France)=Paris}. 
Predicates represent whether a relation holds among two objects and return a boolean value.
For example, if usrA and usrB are friends on Twitter, the predicate \textsc{IsFriend  (usrA,usrB)= true}. 
Predicates and functions can be transformed to each other. 
Given the function \textsc{WifeOf(usrA)=usrB}, the predicate \textsc{IsCouple(usrA,usrB)=true} will naturally hold.
As we will demonstrate later, all functions will be transformed to predicates in graph construction procedure. 

\subsection{Dataset}

We use a random sample of Twitter users---after discarding users with less than 10 tweets--
consisting of 0.5 million Twitter users.
We crawled their published tweets and their network
using the Twitter API\footnote{Due to API limitations, we can crawl at most 2,000 tweets for each user.}, 
resulting in a dataset of roughly 75 million tweets.

\subsection{User Attributes}

In the next sections we first briefly describe how we extract predicates for user attributes (location, education, gender) and
user relations (friend, spouse), and then focus in detail on the extraction
of user preferences (like/dislike).

\subsubsection{Location}\label{sec:location}

Our goal is to associate one of the 50 states of the United States with each user.
While there has been a significant amount of work on inferring the location of a given published tweet 
(e.g., \cite{cheng2010you,davis2011inferring,sadilek2012finding}), there is less  focus on user-level inference. 
In this paper, we employ a rule-based approach for user-location identification. 
We selected out all geo-tagged tweets from a specific user, 
and say an entity $e$ corresponds to the location of current user $i$ if it satisfies
the following criteria, designed to ensure high-precision (although with a natural corresponding
drop in recall):
\begin{enumerate}
\item user $i$ published more than 10 tweets from location $e$ 
\item user $i$ published from location $e$ in at least three different months of a year.
\end{enumerate}
We only consider locations within the United States and entities are matched to state names
via Google Geocoding.  In the end, we are able to extract locations for $1.1\%$ of the users from our dataset.

\subsubsection{Education/Job}

Job and education attributes are extracted by combining a rule based approach 
with an existing probabilistic system described in \cite{li2014weakly}.

First, for each user, we obtained his or her full name and fed it 
into a Google+ API\footnote{\url{https://developers.google.com/+/api/}}.
Many Google+ profiles are publicly accessible and 
many users explicitly list attributes such as their education and job. 
The major challenge involved here is name disambiguation, 
to match users' Twitter accounts to Google+ 
accounts.\footnote{A small property of Google+ accounts contain direct Twitter links. In those cases, accounts can be directly matched.}
We adopted the $friend-shared$ strategy taken in \cite{li2014weakly} that if more than 10 percent of and at least 20 friends are shared by Google+ circles and Twitter followers, we assume that the two accounts point to the same person. $0.8$ percent of users' job or education attributes are finalized based on their Google+ accounts.

For cases where user names can not be disambiguated or no relevant information is obtained from Google+, 
we turn to the system provided by Li et al. \cite{li2014weakly} (for details about algorithms in \cite{li2014weakly}, see Section 6).
This system extracts education or job entities from the published Twitter content of the user,
For each Twitter user, the system returns the education or job entity mentioned in the users Tweets,
associated with a corresponding probability, for example,
$$\textsc{work-in-google (user)}=0.6$$

Since the Li et al. system system requires the user to explicitly mention their education or job entities in their published content,
it is again low-recall: another $0.5$ percent of users' job or education attributes are inferred from the system.

\subsubsection{Gender}

Many frameworks have been devoted to gender prediction from Twitter posts (e.g.,
\cite{burger2011discriminating,ciot2013gender,pennacchiotti2011machine,tang2011s})
studying whether high level tweet features (e.g., link, mention, hashtag
frequency) can help in the absence of highly-predictive user name information.
Since our goal is not guessing the gender without names but rather
studying the extent to which global probabilistic logical inference
over the social network can improve the accuracy of local predictors,
we implement a high-precision rule based approach that uses the
national Social Security Gender Database
(SSGD)\footnote{\url{http://www.ssa.gov/oact/babynames/names.zip}}. 
SSGD contains first-name records annotated for gender for every US birth since
1880 A.D.   Many names are highly gender-specific, while others are ambiguous.
We assign a user a gender if his/her first name appears in the dataset for one gender at least 20 times as often
as for the other.  Using this rule  we assign gender to $78\%$ of the users in our dataset.

\subsection{User Relations}

The user-user relations we consider in this work include \textsc{friend (usrA, usrB)}, 
\textsc{spouse (usrA, usrB)} and \textsc{LiveInSamePlace (usrA, usrB)}.

\noindent {\bf Friend}: Twitter supports two types of following patterns, \textsc{following} and \textsc{followed}. 
We consider two people as friends if they both follow each other (i.e. bidirectional following).
Thus if relation \textsc{Friend(usrA,usrB)} holds, usrA has to be both following and followed by usrB. 
The friend relation is extracted straightforwardly from the Twitter network. 

\noindent {\bf Spouse/Boyfriend/Girlfriend}: For the spouse relation, we again turn to Li et al.'s system \cite{li2014weakly}.
For any two given Twitter users and their published contents, the system returns a score $S_{spouse}$ in the range of [0,1] indicating how likely \textsc{Spouse(usr1,usr2)} relation is to hold. We use a threshold of 0.5 and then for any pair of users with a higher score than 0.5, 
we use a continuous variable to denote the confidence,
the value of which is computed by linearly projecting $S_{spouse}$ into [0,1] space.

\noindent {\bf LiveInSamePlace}: Straightforwardly inferred from the location extraction approach described in \ref{sec:location}.

\subsection{User Preferences: Like and Dislike}

We now turn to user preferences and attitudes---a central focus of our work---and specifically the predicates
\textsc{like(usr,entity)} and \textsc{dislike(usr,entity)}.
Like the large literature on sentiment analysis from social media 
(e.g., \cite{agarwal2011sentiment,kouloumpis2011twitter,pak2010twitter,saif2012semantic}).
our goal is to extract sentiment, but in addition to extract the target or object of the sentiment.
Our work thus resembles other work on sentiment target extraction
(\cite{choi2005identifying,kim2006extracting,yang2013joint})
using supervised classifiers or sequence models based on manually-labeled datasets.
Unfortunately, manually collecting training data in this task
is problematic because (1) tweets talking about what the user \textsc{likes/dislikes} are very sparsely distributed among
the massive number of topics people discuss on Twitter  and 
(2) tweets expressing what the user \textsc{likes} exist in a great variety of scenarios and forms.

To deal with data sparsity issues, we collect training data by combining {\em semi-supervised information harvesting} techniques \cite{davidov2007fully,kozareva2010learning,kozareva2010not,limajor} and the concept of {\em distant supervision} \cite{craven1999constructing,go2009twitter,mintz2009distant} as follows:

{\bf Semi-supervised information harvesting}: 
We applied the standard seed-based information-extraction method of
obtaining training data recursively by 
using seed examples to extract patterns, which are used to 
harvest new examples, which are further used as new seeds to train new patterns.
We begin with pattern seeds including ``I $\_~\_$ like/love/enjoy (entity)", ``I $\_~\_$ hate/dislike (entity)", ``(I think) (entity) is good/ terrific/ cool/ awesome/ fantastic", 
``(I think) (entity) is bad/terrible/awful suck/sucks". 
Entities extracted here should be nouns, which is determined by a Twitter-tuned POS package \cite{owoputi2013improved}.

Based on the harvested examples from each iteration, we train 3 machine learning classifiers: 
\begin{itemize}
\item A tweet-level SVM classifier (tweet-model 1) to distinguish between tweets that 
intend to express like/dislike properties and tweets for all other purposes. 
\item A tweet-level SVM classifier (tweet-model 2) to distinguish between like and 
dislike\footnote{We also investigated a 3-class classifier for
like, dislike and not-related, but found the performance constantly underperforms using separate classifiers.}. 
\item A token-level CRF sequence model (entity-model) to identify entities that are
the target of the users like/dislike.
\end{itemize}

The SVM classifiers are trained using the SVM$_{light}$ package \cite{joachims1999making} with the following features:
\begin{itemize}
\item Unigram, bigram features with corresponding part-of-speech tags and NER labels.
\item Dictionary-derived features based on a subjectivity lexicon \cite{wiebe2005annotating}.
\end{itemize}
The CRF model \cite{lafferty2001conditional} is trained using the CRF++ package\footnote{\url{https://code.google.com/p/crfpp/}} based on the following features:
\begin{itemize}
\item Current word, context words within a window of 3 words and their part-of-speech tags.
\item Name entity tags and corresponding POS tags.
\item Capitalization and word shape. 
\end{itemize}

Trained models are used to harvest more examples, which are further
used to train updated models. We do this iteratively until the stopping condition is satisfied.

{\bf Distant Supervision}: The main idea of distant supervision is
to obtain labeled data by drawing on some external sort of evidence. The
evidence may come from a database\footnote{For example, if datasets
says relation \textsc{IsCapital} holds between Britain and London,
then all sentences with mention of ``Britain" and ``London" are
treated as expressing \textsc{IsCapital} relation
\cite{mintz2009distant,ritter2013modeling}.} or common-sense
knowledge\footnote{Tweets with happy emoticons such as :-)  : ) are
of positive sentiment \cite{go2009twitter}.}.  In this work, we
assume that if a relation \textsc{Like(usr, entity)} holds for a
specific user, then many of their published tweets mentioning the
\textsc{entity} also express the \textsc{Like} relationship and are
therefore treated as positive training data.  
Since semi-supervised approaches heavily rely on seed quality 
\cite{kozareva2010not} and the patterns derived by the recursive framework may be strongly influenced by
the starting seeds, adding in examples from distant supervision 
helps increase the diversity of positive training examples.

An overview of our algorithm 
showing how the {\em semi-supervised approach} is combined with {\em distant supervision} is illustrated in Figure \ref{fig1}.

\begin{figure}[ht]
\rule{8cm}{0.03cm}
{\bf Begin}\\
Train tweet classification model (SVM) and entity labeling model (CRF) based on positive/negative data harvested from 
starting seeds.\\
{\bf While stopping condition not satisfied}:
\begin{tightenu}
\item Run classification model and labeling model on raw tweets. 
Add newly harvested positive tweets and entities to the positive dataset.
\item For any user $usr$ and entity $entity$, if relation \textsc{like(usr,entity)} holds,
add all posts published by $usr$ mentioning $entity$ to positive training data. 
\end{tightenu}
{\bf End}\\
\rule{8cm}{0.03cm}
\caption{Algorithm for training data harvesting for extraction user \textsc{like/dislike} preferences.}
\label{fig1}
\end{figure}

{\bf Stopping Condition}: To decide the optimum number of steps
for the algorithm to stop,
we manually labeled a dataset which contains 200 positive tweets (100 like and 100 dislike) with entities.,
selected from the original raw tweet dataset rather than the automatically harvested data. 
The positive dataset is matched with 800 negative tweets.
For each iteration of data harvesting, we evaluate the performance of the classification models and 
labeling model on this human-labeled dataset, which can be viewed as a development set for parameter tuning. 
Results are reported in Table~\ref{table3}.
As can be seen, the precision score decreases as the algorithm iterates, but the recall rises. 
The best F1 score is obtained at the end of third round of iteration. 

\begin{table}[h]
\centering
\begin{tabular}{|c|c|l|l|l|}
\hline
\multicolumn{1}{|l|}{}                             & \multicolumn{1}{l|}{} & Pre                       & Rec                       & F1 \\ \hline
\multirow{3}{*}{iteration 1}                       & tweet-model 1         & \multicolumn{1}{c|}{0.86} & \multicolumn{1}{c|}{0.40} &0.55    \\ \cline{2-5} 
                                                   & tweet-model 2         & \multicolumn{1}{c|}{0.87} & \multicolumn{1}{c|}{0.84} & 0.85   \\ \cline{2-5} 
                                                   & entity label          & \multicolumn{1}{c|}{0.83} & \multicolumn{1}{c|}{0.40} &0.54    \\ \hline
\multicolumn{1}{|l|}{\multirow{3}{*}{iteration 2}} & tweet-model 1         & 0.78                      & 0.57                      &0.66    \\ \cline{2-5} 
\multicolumn{1}{|l|}{}                             & tweet-model 2         & 0.83                      &  0.86                         & 0.84   \\ \cline{2-5} 
\multicolumn{1}{|l|}{}                             & entity label          & 0.79                      & 0.60                      & 0.68   \\ \hline
\multirow{3}{*}{iteration 3}                       & tweet-model 1         & 0.76                      & 0.72                      &0.74    \\ \cline{2-5} 
                                                   & tweet-model 2         & 0.87                      & 0.86                      & 0.86   \\ \cline{2-5} 
                                                   & entity label          & 0.77                      & 0.72                      &0.74    \\ \hline
\multirow{3}{*}{iteration 4}                       & tweet-model 1         & 0.72                      & 0.74                      &0.73    \\ \cline{2-5} 
                                                   & tweet-model 2         & 0.82                      & 0.82                      & 0.82   \\ \cline{2-5} 
                                                   & entity label          & 0.74                      & 0.70                    & 0.72   \\ \hline
\end{tabular}
\caption{Performance on the manually-labeled devset at different iterations of data harvesting.}
\label{table3}
\end{table}

For evaluation purposes, data harvesting without distant supervision (\textsc{no-distant}) naturally constitutes a baseline. Another baseline we employ is to train a one-step CRF model which directly decide whether a specific token corresponds to a \textsc{like/dislike} entity rather than making tweet-level decision first. 
Both (\textsc{no-distant}) and \textsc{one-step-crf} rely on our recursive framework and tune the number of iterations on the aforementioned gold standards. 
Our testing dataset is comprised of
additional 100 like/dislike property related tweets (50 like and 50 dislike) with entity labels, which are then matched with 400 negative tweets.
The last baseline we employ is the rule based extraction approach by using the seed patterns.
We report the best performance model on the end-to-end entity extraction precision and recall.
To note, end-to-end evaluation setting here is different from what it is in Table 1, 
as if tweet level models make erroneous decision, labels assigned at entity level would be treated as wrong,

\begin{table}[h]
\centering
\begin{tabular}{|c|c|c|c|}\hline
Model&Pre&Rec&F1\\\hline
semi+distant&0.73&0.64&0.682\\\hline
no-distant&0.70&0.65&0.674\\\hline
one-step (CRF)&0.67&0.64&0.655\\\hline
rule&0.80&0.30&0.436\\\hline
\end{tabular}
\caption{Performances of different models on extraction of user preferences (like/dislike) toward entities.}
\label{tab2}
\end{table}

As can be seen from Table \ref{tab2}, about three points of performance boost are obtained by incorporating user-entity information from distant supervision. 
Modeling tweet-level and entity-level information separately yields better performance than incorporating them in a unified model (\textsc{one-step-crf}). 

We apply the model trained in this subsection to our tweet corpora. 
We filter out entities that appear less than 20 times, resulting in roughly 40,000 different entities\footnote{Consecutive entities with same type of NER labels are merged.} in the dataset.

{\bf Entity Clustering}: We further cluster the extracted entities into different groups, 
with an goal of answering questions like 
`if usr1 likes films, how likely would she like the film {\em Titanic}?'

Towards this goal, we train a skip-gram neural language model \cite{mikolov2010recurrent,mikolov2011extensions} based on the tweet dataset using word2vec where each word is represented as a real-valued, low-dimensional vector\footnote{Word embedding dimension is set to 200}.
Skip-gram language models 
draw on local context in order to learn similar embeddings for semantically similar words. 
Next we run a k-means clustering algorithm (k=20) on the extracted entities, using L2 distance.
From the learned clusters, we manually selected out 12 sensible ones,
including food, sports, TV and movies, politics, electronic
products, albums/concerts/songs, travels, books, fashions, financial
stuff, and pets/animals.  Each of the identified clusters is then matched
with a human label.

{\bf Like Attribute Extracted from Network}:

We extract more \textsc{like/dislike} preferences
by using the \textsc{following} network of Twitter. If a
twitter user $e_i$ is followed by current user $e_j$, but not
bidirectionally, and that $e_i$ contains more than 100,000 followers,
we treat $e_i$ as a public figure/celebrity that is liked by current user $e_j$.

\section{Logic Networks}
In this section, we describe MLN and PSL, which have been
widely applied in relational learning and logic reasoning.

\subsection{Markov Logic}
Markov Logic \cite{richardson2006markov} is a probabilistic logic framework which encodes weighted first-order logic formulas
in a Markov network. By translating to logic, the expression 
{\em people from Illinois like the NFL football team Chicago Bears} can be expressed as:
\begin{equation}
\forall x \textsc{live-in}(x,Illinois)\Rightarrow \textsc{like}(x,Chicago Bears) 
\label{equ1}
\end{equation}
Real world predicates are first converted to symbols using logical connectives and quantifiers.
In MLN, each of the predicates (e.g., \textsc{LiveIn} and \textsc{like}) corresponds to a node 
and each formula
is associated with a weighted value $w_i$. The frameworks optimizes the following probability:
\begin{equation}
P(X)=\frac{1}{Z}\prod_i \phi(x_i)^{n_i(x)}
\label{equ2}
\end{equation}
where $\phi(x_i)=\exp(w_i)$.
$Z$ denotes the normalization factor and $x_i$ denotes the states of nodes in the network. In our early example, $x$ could take the following 4 values, i.e., $(l_1, l_2)$, $(\neg l_1, l_2)$, $(l_1, \neg l_2)$ and $(\neg l_1, \neg l_2)$.
$n_i(x)$ is the number of true groundings for state $x_i$.
Consider the simple logic network shown in Equ. \ref{equ1} with weight $w$, given the logic rule that
$f_1 \Rightarrow f_2$ is true iff $f_1$ is false or $f_2$ is true,
we have $P(l_1,\neg l_2)=1/(1+3\exp(w))$ and 
the probability
of each of the other three $\exp(w)/(1+3\exp(w))$.

For inference, the probability of predicate $l_i$ given the rest of the predicates is written as:
\begin{equation}
\begin{aligned}
P(l_i| l_{rest})&=\frac{P(l_i \wedge l_{rest})}{P(l_{rest})}\\
&=\frac{\sum_{x\in l_i\cup l_{rest}}P(x| \cdot)}{\sum_{x\in l_{rest}}P(x| \cdot)}
\end{aligned}
\end{equation}  
Many approaches have been proposed for fast and effective learning for MLNs \cite{lowd2007efficient,niu2011tuffy,singla2005discriminative}.
In this work, we use the discriminative training approach \cite{singla2005discriminative}, 
as will be demonstrated in Section 4.1.

\subsection{Probabilistic Soft Logic}
PSL \cite{beltagy2014probabilistic,kimmig2012short} is another sort of logic reasoning architecture.
It first associates each predicate $l$ with a  soft truth value $I(l)$. 
Based on such soft truth values, PSL 
performs logical conjunction and disjunction in the following ways: 
\begin{equation}
\begin{aligned}
&I(l_1\vee l_2)=max\{0, I(l_1)+I(l_2)-1\}\\
&I(l_1\wedge l_2)=min\{1, I(l_1)+I(l_2)\}\\
\end{aligned}
\end{equation}
Next a given formula $l_1\Rightarrow l_2$ is said to be satisfied if $I(l_1)\leq I(l_2)$. 
PSL defines a variable $d(r)$, the `distance to satisfaction', to capture how far rule r is from being true.
$d(r)$ is given by $max\{0, I(l_2)-I(l_1)\}$. 
For example, if
\begin{equation}
I(\textsc{Spouse(usr1,usr2)})=1 
\end{equation}
and
\begin{equation}
I(\textsc{like(usr1,entity1})=0.6
\end{equation}
then 
\begin{equation}
\begin{aligned}
I(\textsc{Spouse(usr1,usr2)}&\wedge \textsc{like(usr1,entity1}))\\
& = max(0, 1+0.6-1)=0.6
\end{aligned}
\end{equation}
PSL is optimized through  maximizing observed rules in terms of 
distant $d(r)$:
\begin{equation}
\begin{aligned}
&P(I)=\frac{1}{Z}\exp[-\sum_r \lambda_r(d(I))]\\
&Z= \int_I \frac{1}{Z}\exp[-\sum_r \lambda_r(d(I))]
\end{aligned}
\end{equation}
where $Z$ denotes the normalization factor, and $\lambda_r$ denotes the weight for formula. 
Inference can be straightforwardly performed by calculating the distance $d$ between the predicates. 
Compared with MLN, the PSL framework can be efficiently optimized based on a linear  program.
Another key distinguishing feature for PSL is that it uses continuous variables (soft truth values) rather than binary ones in MLN.

\section{Logic Reasoning on Social Networks}
Based on our extraction algorithm in Section 2,
each user $i$, is associated with
a list of attributes and preferences,
and is related by various relations to other users in  a network.
Function symbols are transformed to predicates for graph construction,
where all the nodes in the graph take on binary values (i.e., true or false). 

\subsection{Assumptions and Simplifications}

As existing algorithms might be difficult to scale up to the size of users and attributes we consider, 
we make some assumptions to enable faster learning and inference:

{\bf Cut off Edges}: 
If 
relations
\textsc{like (usrA, entity1)}, \textsc{like (usrB, entity2)} and \textsc{friend (usrA, usrB)} hold, 
but \textsc{entitiy1} and \textsc{entity2} are from different like-entity categories,  
we would say
\textsc{like (usr1, entity1)} and \textsc{friend (usr1, usr2)} 
are independent, which means there would be no edge connecting nodes \textsc{like (entity1)} and \textsc{like (entity2)}
in the Markov network.
As an example, if usrA likes fish and usrB likes football, as fish and football belong to different entity categories, we would treat these two predicates as independent. 

{\bf Discriminative Training for MLN}: We use the approach described in \cite{singla2005discriminative} 
where we assume that we have a priori knowledge about
which predicates
will be evidence and which ones will be queried. Instead of optimizing over all nodes along the graph, the 
system optimizes the probability of predicting the queried nodes given evidence nodes.
This prunes a large number of branches.
Let $Y$ be the set of queried models and $X$ be evidence nodes, 
the system optimizes the conditional probability as follows: 
\begin{equation}
p(Y|X)=\frac{1}{Z}\exp (\sum_{i\in F_{Y}}w_i n_i(x,y))
\end{equation} 
where $F_Y$ denotes all cliques with at least one
node involving a query node.

\subsection{Modeling Missing Values}

A major challenge  is missing values,  for example in situations where users do not mention an entity;
a user not mentioning one entity does not necessarily mean they do not like it. 
Consider the following situation: 
\begin{equation}
\begin{aligned}
\textsc{friend(A,B)} \wedge &\textsc{like(A,soccer)} \\
&~~~~~~~~~~~~~~~~\Rightarrow \textsc{like(B,soccer)}
\end{aligned}
\end{equation}

Drawing this inference in this way is requires that
(1) usrB indeed likes soccer (2) usrB  explicitly mentions soccer in his or her posts. 
Satisfying both premises (especially the latter one) is a luxury. 
Inspired by common existing approaches to deal with missing data \cite{lauritzen1995algorithm,little2002statistical},
we treat users' \textsc{like/dislike} preferences as latent variables,
while what is observed is whether users explicitly mention their preferences in their posts. 
The latent variables and observed variables are connected via a binary distribution 
parameterized by a [0,1] variable $S_{entity}$, indicating how likely a user 
would be to report the correspondent entity in their posts. 

\begin{figure}
\centering
\includegraphics[width=3.5in]{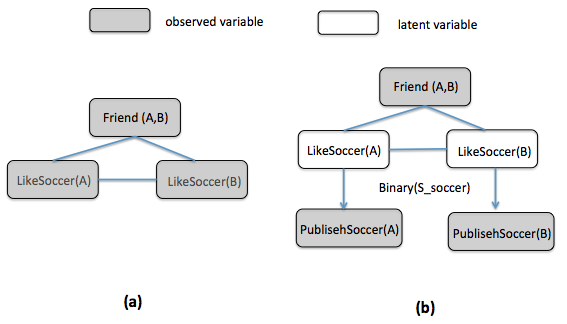}
\caption{(a) Standard Approach (b) Revised version with missing values in MLN.}\label{fig2}
\end{figure}

For MLN, a brief illustration is shown in Figure \ref{fig2}. The conditional probability can be expressed by summing over latent variables. The system can be optimized by incorporating a form of EM algorithm into MLN \cite{singla2006entity}.  

For PSI, each entity is associated with an additional predicate \textsc{mention (usr, entity)}, denoting the situation where any given user 
publishes posts about one specific entity. Predicate \textsc{publish-entity(usr)} comes with the following constraints :
\begin{itemize}
\item $\neg$ \textsc{like-entity(usr)}$\wedge$\textsc{publish-entity(usr)}=0
\item $\neg$ \textsc{dislike-entity(usr)}$\wedge$\textsc{publish-entity(usr)}=0
\end{itemize}
which can be interpreted as saying that a user would mention his like or dislike towards an entity only if he likes or dislikes it.

\subsection{Inference}
Inference is performed on two settings:
\textsc{friend-observed} and \textsc{neigh-latent}\footnote{We draw on a similar idea in \cite{limajor}.}. 
\textsc{friend-observed} addresses the leave-one-out testing to infer one specific attribute or relation 
given all the rest. \textsc{friend-latent} refers to a another scenario where some of the attributes  (or other information)
for multiple users
along the network are missing and require joint inference over multiple values along the graph. 
Real world applications, where network
information can be partly retrieved, likely fall in between.

Inference for  the \textsc{friend-observed}  setting is performed directly from the standard MLN and PSL inference framework,
which is implemented using MCMC for MLN and 
MPE (Most Probable Explanation) for PSL. 
For the \textsc{friend-latent} setting, we need to jointly infer location attributes along the users. 
As the objective function for joint inference would be difficult to optimize (especially since inference on MLN is hard) 
and existing algorithms may not able to scale up to the size of network we consider,
we turn to a greedy 
approach inspired by recent work \cite{li2014weakly,raghunathan2010multi}: attributes 
are initialized from the logic network based on given attributes where missing values are not considered.
Then for each user along the network, we iteratively re-estimate their attributes given the evidence both from
her own attribute values and her friends by performing standard inference in MLN or PSL.
In this way, highly confident predictions will be made based on individual features in the first round, then 
user-user relations would either support or contradict these decisions. We run 3 rounds of iterations. We expect
\textsc{friend-observed} to yield better results than the \textsc{friend-latent} setting since the former benefits from 
gold network information \cite{limajor}.

\section{Experiments}

We now turn to our experiments on using global inference across the logic networks 
to augment the individual local detectors to infer user attributes, user
relations and finally user preferences.
These results are based on the datasets extracted in the previous
section, where each user is represented with a series of extracted attribute values (e.g., like/dislike, location, gender) and users are connected along the social network.  We use $90\%$ of the data as training corpus, reserving $10\%$ for testing, from which
we respectively extract testing data for each relations, attribute, or preference, as described below.

In each case, our goal is to understand whether global probabilistic logical inference over the entire
social network graph improves over baseline classifiers like SVNs that use only local features.

\subsection{User Attributes: Location}

The goal of location inference is to identify the US state the user tweets from, out of the 50 states.
Evaluation is performed on the subset of users for which our rule-based approach in Section 2 identified
a gold-standard location with high precision. We report on two settings.
The \textsc{friend-latent} setting makes joint predictions for user locations across the network while
the more precise \textsc{friend-observed} setting predicts the locations of each user given all other attributes, relations, and preferences.
Baselines we employ include:
\begin{itemize}
\item {\bf Random}: Assign location attributed from distribution based on population\footnote{\url{http://en.wikipedia.org/wiki/List_of_U.S._states_and_territories_by_population}}. 
\item {\bf Unified}: Assign the most populated state in USA (California) to each user.
\item {\bf SVM} and {\bf Naive Bayes}: Train multi-class classifiers where features are the predicted extracted attributes and network information. 
Features we consider include individual information and network information. 
The former encodes the presence/absence of the entities that a user likes or dislikes, and job/education attributes (if information is included in the dataset). 
The latter includes 
\begin{itemize}
\item  The proportion of friends that take one specific value for each attribute. Consider the attribute \textsc{LiveIn-illinois}, feature value for SVM is calculated as follows\footnote{For probabilistic attributes (e.g., education), values are leveraged by weights. }:
$$\frac{\sum_{\text{usr}} {\bf I}(\textsc{LiveIn-illinois(usr))} }{\sum_{\text{usr}} {\bf I}}$$
\item The presence/absence of spouse attribute (if spouse user is identified). 
 \end{itemize}
\item {\bf Only-network}: A simplified version of the model which only relies on relations along the network.
\item {\bf Only-like}: A simplified version of the model which only relies on individual attributes.
\end{itemize}

\begin{table}[h]
\small
\centering
\begin{tabular}{cccc}\hline
Model&Acc&Model&Acc\\\hline
Random&0.093&Unified&0.114\\
SVM&0.268&Naive Bayes&0.280\\
only-network (MLN)&0.272&only-like (MLN)&0.258\\
friend-observed (MLN)&0.342&friend-latent (MLN)&0.298\\
friend-observed (PSL)&0.365&friend-latent (PSL)&0.282\\\hline
\end{tabular}
\caption{Accuracy of different models for predicting location.}
\label{tab5}
\end{table}

\begin{table}
\centering
\begin{tabular}{|c|c|}\hline
Form&Value\\\hline
\multirow{2}{*}{$\frac{\text{Pr}(\textsc{like-ChicagoBear}(A)|\textsc{Live-In-illinois}(A))}{\text{Pr}(\textsc{like-ChicagoBear}(A)|\textsc{Not-Live-In-illinois}(A))}$}&17.8 \\
& \\\hline 
\multirow{2}{*}{$\frac{\text{Pr}(\textsc{like-Barbecue(A)}|\textsc{Live-in-Alabama}(A))}  {\text{Pr}(\textsc{like-Barbecue(A)}|\textsc{NotLive-in-Alabama}(A))}$} &2.8 \\
& \\\hline 
\multirow{2}{*}{$\frac{\text{Pr}(\textsc{like-HOCKEY(A)}|\textsc{Live-in-Minnesota}(A))}  {\text{Pr}(\textsc{like-Barbecue(A)}|\textsc{Live-in-Florida}(A))}$} &7.2 \\
& \\\hline 
\multirow{2}{*}{$\frac{\text{Pr}(\textsc{like-Krumkake(A)}|\textsc{Live-in-North Dakota}(A))}  {\text{Pr}(\textsc{like-Krumkake(A)}|\textsc{NotLive-in-North Dakota}(A))}$} &4.2\\
& \\\hline 
\end{tabular}
\caption{Examples for Locations.}
\label{tab-location}
\end{table}

The performances of the different models are illustrated in Table \ref{tab5}\footnote{This is a 50-class classification problem;accuracy for random assignment without prior knowledge is 0.02$\%$.}.
As expected, \textsc{friend-observed} outperforms \textsc{friend-latent}, detecting locations with an accuracy of about 0.35.
{\bf only-network} and {\bf only-like} models, where evidence is partially considered, consistently underperform settings where evidence is
fully considered. Logic networks, which are capable of capturing the complicated interaction between factors and features, yield better performance than
traditional SVM and Naive Bayes classifiers.

Table \ref{tab-location} gives some examples based on conditional probability calculated from MLN, respectively correspond: 
(1)  people from Illinois like Chicago Bears
(2) People from Alabama like barbecue
(3) People from hockey like hockey
(4) People from North Dakota like Krumkake.

\subsection{User Attributes: Gender}

We evaluate gender based on a dataset of 10,000 users (half male, half female) drawn from the users whose 
gold standard gender was assigned with sufficiently  high precision by the social-security informed system in Section 2.
We only focus on \textsc{neigh-observe} setting.
SVM baseline takes individual and network features as described in Section 5.1.
Table~\ref{TABgender} shows the results.
Using the logic networks across all attributes, relations, and preferences, 
the accuracy of our algorithm is 0.772.
\begin{table}[h]
\centering
\begin{tabular}{|c|c|c|c|}\hline
Model&Pre&Rec&F1 \\\hline
MLN&0.772&0.750&0.761\\\hline
PSL&0.742&0.761&0.751\\\hline
SVM&0.712&0.697&0.704 \\\hline
\end{tabular}
\caption{Performances for Male/Female prediction.}
\label{TABgender}
\end{table}

Of course the performance of the algorithm could very likely be even higher if we were to additionally incorporating 
features designed directly for the gender ID task (such as entities mentioned, links, and especially the wide
variety of writing style features used in work such as \cite{ciot2013gender}, which achieves gender ID accuracies of 0.85 on a different dataset).
Nonetheless, the fact that global probabilistic inference over the network of attributes and relations
achieves such high accuracies without any such features points to the strength of the network approach.

Table \ref{tab-gender} gives some examples about gender preference inferred from MLN. As can be seen, males prefer sports while females prefer fashions (as expected). Females emphasize more on food and movies than males, but not significantly.

\begin{table}
\centering
\begin{tabular}{|c|c|}\hline
Form&Value\\\hline
\multirow{2}{*}{$\frac{\text{Pr}(\textsc{like-Fashion(A)}|\textsc{IsFemale}(A))}  {\text{Pr}(\textsc{like-Fashion(A)}|\textsc{IsMale}(A))}$} &16.9\\
& \\\hline 
\multirow{2}{*}{$\frac{\text{Pr}(\textsc{like-Sports(A)}|\textsc{IsMale}(A))}  {\text{Pr}(\textsc{like-Sports(A)}|\textsc{IsFeMale}(A))}$}&18.0 \\
& \\\hline 
\multirow{2}{*}{$\frac{\text{Pr}(\textsc{like-Food(A)}|\textsc{IsFemale}(A))}  {\text{Pr}(\textsc{like-Food(A)}|\textsc{IsMale}(A))}$}&2.1 \\
& \\\hline 
\multirow{2}{*}{$\frac{\text{Pr}(\textsc{like-movies(A)}|\textsc{IsFemale}(A))}  {\text{Pr}(\textsc{like-movies(A)}|\textsc{IsMale}(A))}$}&1.6 \\
& \\\hline 
\end{tabular}
\caption{Examples for Genders.}
\label{tab-gender}
\end{table}

\subsection{Predicting Relations Between Users}

We tested relation prediction  on the detection of the three relations defined in section 2:
\textsc{friend}, \textsc{spouse} and \textsc{LiveInSameLocation}.
Positive training data is selected from pairs of users among whom one specific type of relation
holds while random user pairs are used as negative examples. 
We weighted toward negative examples to match the natural distribution
Statistics are shown in Table \ref{tab6}.
\begin{table}[h]
\centering
\begin{tabular}{ccc}\hline
Relation&Positive&Negative\\\hline
Friend&20,000&80,000\\
Spouse&1,000&5,000\\
LiveInSameLocation&5,000&20,000\\\hline
\end{tabular}
\caption{Dataset statistics for relation prediction.}
\label{tab6}
\end{table}

For relation evaluation, we only focus on the \textsc{neigh-observe} setting. 
Decisions are made by comparing the conditional probability that a specific relation holds given other types of information, for example
 Pr(\textsc{Spouse(A,B)}|{\bf $\cdot$}) and 1-Pr(\textsc{Spouse(A,B)}|{\bf $\cdot$}).
Baselines we employ include:
 
\begin{itemize}
\item SVM: We use co-occurrence  of attributes as features: 
$$\textsc{Like-Entity1(A)}\wedge\textsc{Like-Entity2(B)}$$
For \textsc{LiveInSameLocation} prediction, the location identification classifier without any global information naturally constitutes a baseline,
where two users are viewed as living in the same location if classifiers trained in 5.2 assigned them the same location labels. 
\item Random: Assign labels randomly based on the proportion of positive examples. We report the theoretical values of precision and recall, which are given by:
$$\text{Pre, Rec}=\frac{\#\text{ positive-examples}}{\#\text{total-examples} }$$
\end{itemize}

The performance of
the  different approaches are reported in Table \ref{tab6}. 
As can be seen, the logic models consistently yield better performances than SVMs on relation prediction tasks 
due to their power in leveraging the global interactions between different sources of evidence.

\begin{table}
\centering
\begin{tabular}{|c|c|c|c|c|}
\hline
Relation                                                                       & Model                                                    & Pre   & Rec   & F1 \\ \hline
\multirow{4}{*}{Friend}                                                        & MLN                                                      & 0.580 & 0.829 & 0.682   \\ \cline{2-5} 
                                                                               & PSL                                                      & 0.531 & 0.850 & 0.653   \\ \cline{2-5} 
                                                                               & SVM                                                      & 0.401 & 0.745 & 0.521   \\ \cline{2-5} 
                                                                               & Random                                                   & 0.200 & 0.200 & 0.2   \\ \hline
\multirow{4}{*}{Spouse}                                                        & MLN                                                      & 0.680 & 0.632 &0.655    \\ \cline{2-5} 
                                                                               & PSL                                                      & 0.577 & 0.740 &0.648    \\ \cline{2-5} 
                                                                               & SVM                                                      & 0.489 & 0.600 & 0.539   \\ \cline{2-5} 
                                                                               & Random                                                   & 0.167 & 0.167 &  0.167  \\ \hline
\multirow{5}{*}{\begin{tabular}[c]{@{}c@{}}LiveInSame\\ Location\end{tabular}} & MLN                                                      & 0.592 & 0.704 &0.643    \\ \cline{2-5} 
                                                                               & PSL                                                      & 0.650 & 0.741 &0.692    \\ \cline{2-5} 
                                                                               & SVM                                                      & 0.550 & 0.721 &  0.624  \\ \cline{2-5} 
                                                                               & \begin{tabular}[c]{@{}c@{}}SVM\\ (location)\end{tabular} & 0.504 & 0.695 &0.584    \\ \cline{2-5} 
                                                                               & Random                                                   & 0.200 & 0.200 &0.200    \\ \hline
\end{tabular}
\caption{Performances for Relation Prediction. Performances about random are theoretical results.}
\label{tab6}
\end{table}

\subsection{Predicing Preference: Likes or Dislikes}

Evaluating our ability to detect user preferences is complex since, as mentioned in Section 4.2,
we don't  gold-standard labels of Twitter users' true attitudes towards different entities.
That is, we don't actually know what users actually like: only what they {\em say} they like.
We therefore evaluate our ability to detect what users say about an entity.

We evaluate two distinct tasks, beginning with the simpler:  given that the user
talked about an entity, was their opinion positive or negative. 

We then proceed to the much more difficult task of predicting whether a user
will talk about an entity at all, and if so whether her opinion will be positive or negative.

In both tasks our task is to estimate without using the text of the message.
This is because our goal is to understand how useful the social network structure  is alone
in solving the problem.  This information could then easily be combined with standard
sentiment-analysis techniques in future work.

Evaluations are performed under both the \textsc{friend-observed} setting and the \textsc{friend-latent} setting. 

\subsubsection{Predicting Like/Dislike}

We begin with the  scenario in which we know that an entity $e$ is already mentioned by user $i$
and we try to predict a user's attribute towards $e$ without looking at text-level evidence.
The goal of this  experiment is to  predict sentiment
(e.g., whether one likes Barack Obama) given other types of attributes
of the user himself (e.g., where he lives) or his network (e.g.,
whether his friends hate Mitt Romney) but without using sentiment-analysis features of the text itself.

We created a test set in which
the like/dislike preferences are expressed toward
entities (extracted in Section 2) that are frequent from our database, which has a total of 92 distinguished entities (e.g., BarackObama, New York Knicks). We extracted 
1000 like examples and 1000 dislike examples (e.g., \textsc{like-BarackObama(user)}, \textsc{dislike-BarackObama(user)}).

Predictions are make by comparing Pr( \textsc{like-entity (user,entity)}|{\bf $\cdot$}) and Pr( \textsc{dislike-entity (user,entity)}| {\bf $\cdot$}).
We extracted gold standards for each data point, with 0.5 random guess accuracy.
Evaluations are performed in terms of prediction and recall.
We only consider the \textsc{neigh-latent} setting.

The Baselines we employ include: 
\begin{itemize}
\item SVM:  We train binary SVM classifiers to decide,
for a specific entity $e$, whether a user likes/dislikes $e$. 
Features include individual attributes values (e.g., like/dislike, location, gender, etc) and network information (attributes from his friends along the network)
\item {\bf Collaborative Filtering (CF)}: CF \cite{debnath2008feature,goeksel2010system,jamali2010matrix,kautz1997referral} accounts for a popular approach in recommendation system, 
which utilizes the information of the user-item
matrix for recommendations. 
The key idea of CF is to recommend similar items to similar users. 
We view the like/dislike entity prediction as entity recommendation problem
and adopt the approach described in \cite{sarwar2001item} by constructing user-user similarity matrix from weighted cosine similarity calculated from shared attributes and network information. 
Entity-entity similarity is computed based on entity embedding (described in Section 2).
As in \cite{sarwar2001item}, a regression model is trained to fill out $\{0, 1\}$ value in user-entity matrix indicating whether a specific user likes/hates one specific entity.  
Prediction is then made based on a weighted nearest neighbor algorithm. 
\end{itemize}

Results  are reported in Table \ref{TABlike}. As can be seen, MLN and PSL outperform other baselines. 

\begin{table}
\centering
\begin{tabular}{|c|c|c|c|}\hline
Model&Pre&Rec&F1 \\\hline
MLN&0.802&0.764&0.782\\\hline
PSL&0.810&0.772&0.791\\\hline
SVM&0.720&0.737&0.728 \\\hline
CF&0.701&0.694&0.697 \\\hline
\end{tabular}
\caption{Performances for Like/Dislike prediction.}
\label{TABlike}
\end{table}

\subsubsection{Predicting Mentions of Likes/Dislikes}

We are still confronted with the missing value problem of Section 4.2,
where we don't know what users actually believe, so we can only try
to predict what they will say they believe.
But  where the previous section assumed we knew that the user talked about an entity
and just predicted the sentiment, 
we now turn to the much more difficult task of predicting both whether a user
will mention an entity  and what the users attitude toward the entity is.
Evaluations are again  performed under the \textsc{friend-observed} setting and the \textsc{friend-latent} setting. 

We construct the testing dataset using a random sample of 2,000 users with an average number of 3.2
like/dislike entities mentioned per user (a total number of 4,300 distinct entities). Baselines we employ include:
\begin{itemize}
\item {\bf Random}: Estimate the overall popularity of a specific entity being liked by the whole population.
The probability $p_{entity}$ is given by:
$$p_{entity}=\frac{\sum_{usr} {\bf I(\textsc{Like-Entity}(usr)=true)}}{\sum_{usr}{\bf I}}$$
The decision is made by sampling a $\{0,1\}$ variable from a binary distribution parameterized by $p_{entity}$.
\item {\bf SVM} and {\bf Naive Bayes}: We train SVM and Naive Bayes classifiers to decide whether 
a specific user would express his like/dislike attitude towards a specific entity. 
Features include individual attributes values and network information (For feature details, see Section 5.1).
\item {\bf Collaborative Filtering (CF)}: As described in the previous section.
\end{itemize}

Performances are evaluated in terms of precision and recall, reported in Table \ref{tab6}.

Note that predicting like/dislike mention is an extremely difficult task, since users tweet about only 
a very very small percentage of all the entities they like and dislike in the world.
Predicting which ones they will decide to talk about is a difficult task
requiring much more kinds of evidence about the individual and the network that our system
has access to.

Nonetheless, 
given the limited information we have at hand, and considering the great number of entities,
our proposed model does surprisingly well,
with about $7\%$ precision and $12\%$ recall, significantly outperforming
Collaborative Filtering, SVM and Naive Bayes. 
\begin{table}
\small
\centering
\begin{tabular}{cccc}\hline
Model&Pre&Rec&F1\\\hline
Random&<0.001&<0.01&<0.01 \\
SVM&0.023&0.037&0.028\\
Naive Bayes&0.018&0.044&0.025 \\
CF&0.045&0.060&0.051\\
friend-latent (MLN)&0.054&0.066&0.059\\
friend-latent (PSL)&0.067&0.061&0.064\\
friend-observed (MLN)&0.072&0.107&0.086\\
friend-observed (PSL)&0.075&0.120&0.092\\\hline
\end{tabular}
\caption{Performances of different models for like/dislike mention prediction.}
\label{tab6}
\end{table}

\begin{table*}
\centering
\small
\begin{tabular}{|c|c|c|c|}\hline
logic form&probability from MLN& &description\\\hline
\textsc{friend(A,B)}$\wedge$ \textsc{friend(B,C)}$\Rightarrow$ \textsc{friend(A,C)}&0.082& &friends of friends are friends\\\hline
\textsc{couple(A,B)}$\wedge$ \textsc{friend(B,C)}$\Rightarrow$ \textsc{friend(A,C)}&0.127& &one of a couple and the other's friend are friends\\\hline
\textsc{friend(A,B)}$\wedge$ \textsc{lke-sports(A)}$\Rightarrow \textsc{lke-sports(B)}$&0.024& &friend of a sports fan likes sports\\\hline
\textsc{friend(A,B)}$\wedge$ \textsc{lke-food(A)}$\Rightarrow \textsc{lke-food(B)}$&0.018& &friend of a food fan likes food\\\hline
\textsc{friend(A,B)}$\wedge$ \textsc{lke-fashion(A)}$\Rightarrow \textsc{lke-fashion(B)}$&0.030& &friend of a fashion fan likes fashion\\\hline
\textsc{couple(A,B)}$\wedge$ \textsc{lke-sports(A)}$\Rightarrow \textsc{lke-sports(B)}$&0.068& &wife/husband of a sports fan likes sports\\\hline
\textsc{couple(A,B)}$\wedge$ \textsc{lke-food(A)}$\Rightarrow \textsc{lke-food(B)}$&0.085& &wife/husband of a food fan likes food\\\hline
\textsc{friend(A,B)}$\Rightarrow$ \textsc{liveinSameplace(A,B)}&0.160& &friends live in the same location\\\hline
\textsc{couple(A,B)}$\Rightarrow$ \textsc{liveinSameplace(A,B)}&0.632& &couple live in the same location\\\hline
\textsc{Work-In-IT-company(A)}$\Rightarrow$ \textsc{like-electronic-device}&0.242& &people work in IT companies like electronic devices\\\hline
\textsc{Student(A)$\Rightarrow$ \textsc{lke-sports(A)}}&0.160&&student users like sports\\\hline
\textsc{Ivy-Student(A)$\Rightarrow$ \textsc{lke-sports(A)}}&0.125&&student users from Ivy schools like sports\\\hline
\textsc{Spouse(A,B)$\Rightarrow$ \textsc{Friend(A,B)}}&0.740&&couples are friends\\\hline
\end{tabular}
\caption{Examples of inference probability from the proposed system. }
\label{tab8}
\end{table*}

\section{Related Work}
\label{related_work}
This work is related to four different research areas.

{\bf Information Extraction on Social Media} :
Much work has been devoted to automatic extraction of well-structured information profiles from online social media, which mainly fall into two 
major levels: at public level \cite{lin2010pet,lin2011smoothing,wiebe2005annotating} or at user level.
The former includes public event identification \cite{diao2012finding}, event tracking \cite{popescu2011extracting} or event-referring expression extraction \cite{ritter2012open}.
The latter focus on user studies,
examining users' interests \cite{banerjee2009user},
timeline \cite{li2014timeline},
personal events \cite{limajor} or 
individual attributes 
such as 
age \cite{rao2010classifying,rao2010detecting}, gender \cite{ciot2013gender},  political polarity \cite{conover2011political}, locations \cite{sadilek2012finding}, jobs and educations \cite{li2014weakly}, student information (e.g., major, year of matriculation) \cite{mislove2010you}. 

The first step of proposed approach highly relies on attribute extraction algorithm described in \cite{li2014weakly} which extracts three categories of user attributes (i.g., education, job and spouse) for a given user based on their posts.  \cite{li2014weakly} gathers training data based on the concept of distant supervision where Google+ treated used as ``knowledge base" to provide supervision. The algorithm returns the probability of whether the following predicates hold: \textsc{Work-in(usr,entity)} (job), \textsc{Study-at(usr,entity)} (education) and \textsc{Spouse(usr1,usr2)} (spouse).

{\bf Homophily}: Our work is based on the fundamental 
homophily property of online users \cite{mcpherson2001birds}, which assumes that people sharing more
attributes or background have a higher chance of becoming friends in social
media\footnote{summarized by the proverb ``birds of a feather
flock together" \cite{al2012homophily}.}, and that
friends (or couples, or people living in the same location) tend to share more attributes. 
Such properties have been harnessed for applications like community detection \cite{yang2013overlapping} or friend recommendation \cite{guy2010social}. 

{\bf Data Harvesting}: 
The techniques adopted in like/dislike attribute extraction are related to a strand of work in data harvesting/information extraction, 
the point of which is to use some seeds to harvest some data, which is used to learn additional rules or patterns to harvest more data \cite{davidov2007fully,igo2009corpus,kozareva2010learning,kozareva2010not,riloff1999learning}. 
Distant supervision is another methodology for data harvesting \cite{craven1999constructing,hoffmann2011knowledge,mintz2009distant}
that relies on structured data sources as a source of supervision for data harvesting from raw text.

{\bf Logic/Relational Reasoning}: Logic reasoning, usually based on first-order logic representations, can be tracked back to the early days of AI \cite{montague1970universal,robinson1965machine},
and has been adequately explored since then (e.g.,  \cite{berant2013semantic,costa2002clp,goertzel2014probabilistic,jaeger1994probabilistic,kwiatkowski2013scaling,lewis2013combined,richardson2006markov,riedel2013relation,rothfeature,schoenmackers2010learning,wangprogramming2013,wangstarai2014,wangstructurelearning2014,wangMLJ2014}). 
A variety of reasoning models have been proposed, based on ideas or concepts from the fields of graphical models, relational logic, or programming languages \cite{brocheler2012probabilistic,broecheler2010computing,muggleton1994inductive},
each of which has it own generalization capabilities in terms of different types of data.
Frameworks include Stochastic Logic Programs \cite{muggleton1996stochastic} which combines logic programming and log-linear models, 
Probabilistic Relational Networks \cite{friedman1999learning} which incorporates Bayesian networks for reasoning, 
Relational Markov Networks  \cite{taskar2002discriminative} that 
uses dataset queries as cliques and model the state of clique in a Markov network, 
Relational Dependency Networks \cite{neville2007relational} which combines
Bayes networks and Markov networks, and  probabilistic similarity logic \cite{brocheler2012probabilistic} 
which jointly considers probabilistic reasoning about similarities and relational structure.

A great number of applications benefit from logical reasoning, including 
natural language understanding (e.g., \cite{berant2013semantic}), 
health modeling \cite{fakhraeinetwork},  
group modeling \cite{huang2012social}, 
web link based clustering \cite{flake2000efficient}, 
object identification \cite{domingos2004multi}, trust analysis \cite{huang2012probabilistic}, and many more.

\section{Conclusion and Discussion}

In this work, we propose a framework for applying probabilistic logical reasoning
to inference problems on on social networks. 
Our two-step procedure first extracts logical predicates, each associated with a probability,
from social networks, and then performs logical reasoning.
We evaluated our system on predicting user attributes (gender, education, location),
user relations (friend, spouse, same-location), and user preferences (liking or disliking
different entities).
Our results show that using probabilistic  logical reasoning over the network
improves the accuracy of the resulting  predictings, demonstrating the effectiveness of the proposed framework.

Of course the current system is particularly weak in recall, since many
true user attributes or relations are simply never explicitly expressed
on platforms like Twitter.  
Also, the ``gold-standard" first-order logics extracted are not really gold-standard.
One promising perspective is to integrate user information from
different sorts of online social media. Many websites directly
offer gold-standard attributes; Facebook contains
user preference for movies, books, religions, musics or locations;
LinkedIn offers comprehensive professional information. 
Combining these different types of information will offer more evidence for
decision making. 
 

\bibliographystyle{abbrv}
\bibliography{sigproc}  
\end{document}